\documentclass[amsmath,amssymb,aps,reprint,superscriptaddress]{revtex4-1}

\usepackage{graphicx}
\usepackage{dcolumn}
\usepackage{bm}

\makeatletter
\AtBeginDocument{\@ifpackageloaded{natbib}{\ifNAT@numbers\if@filesw\immediate\write\@auxout{\string\global\string\NAT@numberstrue}\fi\fi}{}}
\makeatother \usepackage{amstext}

\begin{document}

\preprint{}

\title{Shot noise induced by nonequilibrium spin accumulation}



\author{Tomonori Arakawa}
\email[]{arakawa@phys.sci.osaka-u.ac.jp}
\affiliation{Department of Physics, Graduate School of Science, Osaka University, 1-1 Machikaneyama, Toyonaka, Osaka 560-0043, Japan}

\author{Junichi Shiogai}
\affiliation{Department of Materials Science, Tohoku University, 980-8579 Sendai, Miyagi, Japan}

\author{Mariusz Ciorga}
\author{Martin Utz}
\author{Dieter Schuh}
\affiliation{Institute of Experimental and Applied Physics, University of Regensburg, D-93040 Regensburg, Germany}
\author{Makoto Kohda}
\affiliation{Department of Materials Science, Tohoku University, 980-8579 Sendai, Miyagi, Japan}
\affiliation{PRESTO, Japan Science and Technology Agency, 332-0012 Kawaguchi, Saitama, Japan}
\author{Junsaku Nitta}
\affiliation{Department of Materials Science, Tohoku University, 980-8579 Sendai, Miyagi, Japan}

\author{Dominique Bougeard}
\author{Dieter Weiss}
\affiliation{Institute of Experimental and Applied Physics, University of Regensburg, D-93040 Regensburg, Germany}

\author{Teruo Ono}
\affiliation{Institute for Chemical Research, Kyoto University, 611-0011 Uji, Kyoto, Japan}

\author{Kensuke Kobayashi}
\email[]{kensuke@phys.sci.osaka-u.ac.jp}
\affiliation{Department of Physics, Graduate School of Science, Osaka University, 1-1 Machikaneyama, Toyonaka, Osaka 560-0043, Japan}



\date{\today}

\begin{abstract}
When electric current passes across a potential barrier, the partition process of electrons at the barrier gives rise to the shot noise, reflecting the discrete nature of electric charge.
Here we report the observation of excess shot noise connected with a spin current which is induced by a non-equilibrium spin accumulation in an all-semiconductor lateral spin valve device.
We find that this excess shot noise is proportional to the spin current.
Additionally, we determine quantitatively the spin-injection induced electron temperature by measuring the current noise.
Our experiments show that spin accumulation driven shot noise provides a novel means to investigate non-equilibrium spin transport.
\end{abstract}

\pacs{}

\maketitle


In 1918, Schottky argued that the electric flow in a vacuum tube fluctuates in a unique way such that the spectral density of the fluctuations is proportional to the unit of charge $e$ ($e>0$) and to the mean current \cite{ref1}. 
This is the shot noise, the direct consequence of the discreteness of the electron charge. 
Now, as an electron possesses not only charge but also spin, one may ask how the discreteness of electron spin affects the current fluctuations. 
Although such spin-dependent shot noise has been discussed theoretically in various contexts \cite{ref2,ref3,ref4,ref5,ref5p5,ref6,ref6p5,ref7,ref8,ref9}, it has never been evidenced experimentally. 

Recently it was pointed out that a non-equilibrium spin accumulation, which can be generated, e.g., by electrical spin injection, modifies the current noise spectrum and allows measuring the magnitude of the spin accumulation electrically \cite{ref9}.
While Meair \textit{et al}. explore in this theoretical work the noise measured between different contacts of a mesoscopic cavity with and without non-equilibrium spin accumulation, we analyze here noise measured across a tunneling barrier where one of the contacts carries a non-equilibrium spin accumulation while the other is in equilibrium (see inset of Fig. 1(a)).
Assuming a linear $I$-$V$ characteristic of the device the current spectral density $S$ is for $\Delta \mu=0$ proportional to the applied voltage $V$ due to shot noise.
For finite spin accumulation, however, $S$ is for $T=0$ given by $S \propto |eV+\Delta \mu/2|+|eV-\Delta \mu/2|$.
This means that at $V = 0$ the noise density is finite.
Plotted as a function of $V$ the spin accumulation appears as a plateau of width $\Delta \mu /e$, shown in Fig. 1(a).
Finite temperatures will smear out the pronounced kinks at $\pm \Delta \mu /2e$.
Using instead of an unpolarized contact in thermal equilibrium a ferromagnetic one offers even more options to measure the spin accumulation electrically, as is pointed out below.
In case of a ferromagnetic contact the expression for $S$ gets modified and reads
\begin{equation}
S_{\uparrow /\downarrow} \propto \alpha |eV\pm \Delta \mu/2|+\beta |eV\mp \Delta \mu/2|,
\label{eq1}
\end{equation}
Here $\alpha$ and $\beta$ ($\alpha>\beta$) define the ratio of the tunnel conductances of majority and minority spins ($\alpha+\beta=1$), and $\uparrow /\downarrow $ denotes the spin direction with respect to the magnetization of the detector contact.
The resulting $S(V)$ relation is plotted in Fig. 1(b).
In addition to the kinks at $\pm \Delta \mu /2e$ the spin-up and spin-down noise density is horizontally shifted by $\gamma  \Delta \mu /e$ where $\gamma $ is the spin asymmetry coefficient given by $\gamma =\alpha -\beta$.
Below we denote the vertical shift of $S_{\uparrow}$ and $S_{\downarrow}$ as excess shot noise, corresponding to the noise which is entirely generated by the spin accumulation.

\begin{figure}[bp]
\center \includegraphics[width=.90\linewidth]{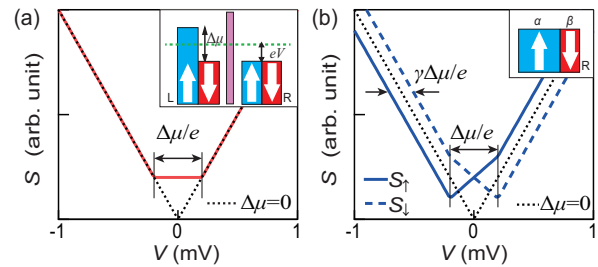}
\caption{(color online). (a) Current noise at $T=0$ for $\Delta \mu =400~\mu $eV, when a non-magnetic detector electrode is used \cite{ref9}. The inset shows the schematic energy diagram. The dotted curve indicates shot noise without spin accumulation. (b) Current noise when a ferromagnetic contact is used on one side of the tunneling barrier as detector electrode ($\alpha =0.75$ and $\beta =0.25 $).}
\label{Fig:1re}
\end{figure}

In this letter we demonstrate the validity of the concepts presented above by measuring shot noise across a tunneling barrier in the presence of spin accumulation.
By using a ferromagnetic detector electrode, we have successfully extracted the relation between the spin current and the corresponding excess shot noise, and found that the estimated Fano factor directly shows that the spin degree of freedom is preserved in the tunneling process.
Given the importance of shot noise in various fields, especially in device technology \cite{ref13} and mesoscopic physics \cite{ref14,ref15,ref16}, excess shot noise due to spin accumulation could not only serve as a unique probe to explore spin-dependent nonequilibrium transport process but also shed new light on the recently emerging field of spin noise spectroscopy \cite{ref17,ref18,ref19, ref20}.

\begin{figure}[bp]
\center \includegraphics[width=.90\linewidth]{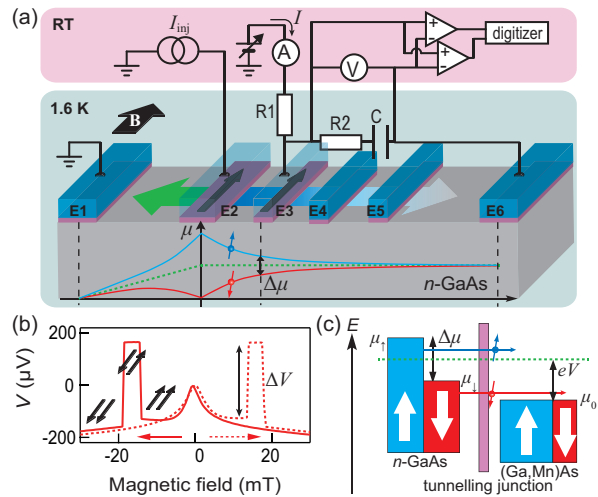}
\caption{(color online). (a) Schematic diagram of the sample and measurement system. Six (Ga,Mn)As electrodes (E1 to E6) are placed on the $n$-GaAs channel, where E2 ($4~\mu \mbox{m} \times 50~\mu \mbox{m}$ size) is an injection electrode, while either E3, E4, or E5 ($0.5~\mu \mbox{m} \times 50~\mu \mbox{m}$ size) is detection electrode. The center-to-center spacing between the neighboring electrodes is $5~\mu $m. Schematic spatial dependence of the each chemical potential in the $n$-GaAs channel is illustrated. (b) Typical non-local voltage signal for $I_{\mathrm{inj}}=-23~\mu $A. A peak observed around zero magnetic field is induced by dynamic nuclear spin polarization (DNP) \cite{ref12}. This effect is irrelevant to the present result, as the noise measurement was performed outside of the DNP region. (c) Schematic energy diagram at the detection electrode in the presence of $eV$ and $\Delta \mu$.}
\label{Fig:1}
\end{figure}

A spin accumulation can be generated, e.g., by electrical spin injection \cite{ref21,ref22}.
To this end, we prepared a lateral all semiconductor spin valve device, which was fabricated from a single epitaxial wafer grown by molecular beam epitaxy on a (001) GaAs substrate, consisting of, in the growth order, GaAs buffer, AlGaAs/GaAs superlattice, $1~\mu $m $n$-GaAs channel, 15 nm GaAs with linearly graded doping $n\to n^{+}$ ($n=2\times 10^{16}~\mathrm{cm}^{-3}$ and $n^{+}=5\times 10^{18}~\mathrm{cm}^{-3}$), 8 nm $n^{+}$-GaAs, 2.2 nm AlGaAs, and 50 nm (Ga$_{0.945}$Mn$_{0.055}$)As.
Due to the high $n^{+}$-doped GaAs region adjacent to the degenerately $p$-doped  (Ga,Mn)As layer, a tunneling Esaki diode structure is formed at the junction, enabling efficient generation and detection of spin accumulation in $n$-GaAs.
The wafer was patterned into $50~\mu $m wide mesas along [110] direction by standard photolithography and wet chemical etching.
Then six ferromagnetic (Ga,Mn)As electrodes (E1 to E6) are defined by electron beam lithography and by wet chemical etching down to the lightly doped $n$-GaAs channel, as schematically shown in Fig. \ref{Fig:1}(a) \cite{ref11,ref12,SM}.
Each ferromagnetic electrode serves a specific purpose; one is used as an injection electrode (E2), three as detection electrodes (E3, E4, and E5), and two as reference electrodes (E1 and E6).
Only the E1 electrode is grounded.

By applying a constant current $I_{\mathrm{inj}}$ to E2 (see Fig. \ref{Fig:1}(a)), we inject spin polarized electrons into the $n$-GaAs channel.
As a consequence, spin accumulation, i.e., splitting of the chemical potentials ($\Delta \mu =\mu _{\uparrow }-\mu _{\downarrow }$) for spin-up and spin-down electrons ($\mu _{\uparrow }$ and $\mu _{\downarrow }$, respectively) occurs underneath the injection contact and diffuses to both sides.
Figure \ref{Fig:1}(b) shows a typical spin accumulation signal obtained  by measuring the non-local DC voltage difference between contacts E3 and E6 as a function of the in-plane magnetic field $B$ (see Fig. \ref{Fig:1}(a)).
All the measurements were carried out at 1.6 K in a variable temperature insert. 
The abrupt voltage changes, displayed in Fig.1(b), correspond to magnetization switching of E2 or E3.
Different coercive fields  were adjusted by different width of the respective contacts.
The voltage change $\Delta V$ between parallel (P) and  antiparallel (AP) magnetisation configurations, being 0.30 mV here, is proportional to $\Delta \mu $ and given by  $\Delta V=\gamma \Delta \mu/e$ \cite{theo1,theo2}.
For our device we extract $\gamma =0.82\pm 0.03$ in the small $I_{\mathrm{inj}}$ limit \cite{ref11,SM}.

To measure $S$ through the tunneling barrier formed by the Esaki diode at the $p$-(Ga,Mn)As/$n$-GaAs interface at the corresponding contact, we use an additional circuit as shown in Fig. \ref{Fig:1}(a).
Three passive components are placed at 1.6~K; two surface mounted metal-film resistors (R1$=1~\mathrm{M}\Omega $, R2$=1~\mathrm{k}\Omega $) and a laminated ceramic capacitor (C$=1~\mu $F).
$S$ is converted to voltage noise by R2, while R1 prevents the current noise from leaking to the voltage source.
The two sets of voltage noise signals are independently amplified by two amplifiers (NF LI-75A) at room temperature, and are recorded at a two-channel digitizer.
Cross-correlation spectra were obtained in the frequency range between 16 kHz and 160 kHz (9001 points) \cite{ref23,ref24}.
To extract $S$ across the tunneling barrier, we carefully calibrated the measurement system and eliminated thermal noise from R2 and the channel resistance \cite{SM}.
We experimentally confirmed that the contribution of frequency-dependent noise, such as $1/f$ noise, does not affect $S$ for the parameter range examined. 

\begin{figure}[bp]
\center \includegraphics[width=.92\linewidth]{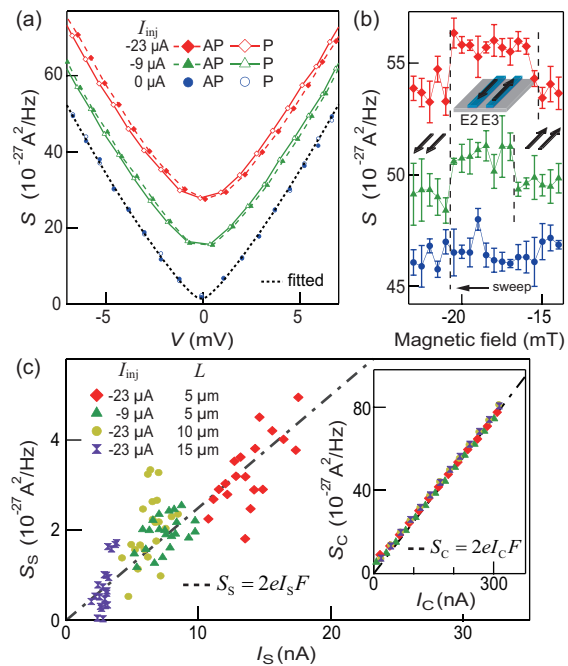}
\caption{(color online). (a) Measured $S$ at E3 as a function of $V$ for P and AP configurations for several injection currents. The curves are offset vertically by $1\times 10^{-26}~\mathrm{A}^{2}/$Hz for clarity. The dotted curve is the fitted curve from Eqn. (1). The error bar for each point is $\pm 0.4\times 10^{-27} \mathrm{A}^{2}/$Hz. (b) $B$-dependence of $S$ measured with keeping $V$ constant ($V=-6.8$~mV) for $I_{\mathrm{inj}}=0, -9$, and $-23~\mu $A (from bottom to top). The thick arrows denote the magnetization directions of E2 and E3. (c) $S_{S}$ versus $I_{S}$ for the bias region ($|eV|>\Delta \mu /2$, $2k_{\mathrm{B}}T$) for several injection currents and for different detection electrodes. The dashed line is the linear relation with $F=0.77$. The inset shows the counterpart of the main graph for $S_{C}$ versus $I_{C}$.}
\label{Fig:2}
\end{figure}

The measurement procedure is as follows \cite{SM}: together with the constant spin injection current $I_{\mathrm{inj}}$, which determines the manitude of the spin accumulation, we inject a small current $I$ to one of the detection electrodes (either E3, E4, or E5) using a constant current technique (see Fig. \ref{Fig:1}(a)).
We directly measure the voltage drop $V$, and $S$ across the tunneling barrier as schematically shown in Fig. \ref{Fig:1}(c).
The dependence of these values on current $I$ was measured for both P and AP configurations.
Note that by reversing the contact magnetization we switch the up and down direction of the accumulated spins in the $n$-GaAs channel.
As $I$ is set between -300 nA and 300 nA, well below $I_{\mathrm{inj}}$, the influence of this probe current on the spin accumulation is negligibly small.
The data shown in Fig. \ref{Fig:2}(a) are obtained for injection currents $I_{\mathrm{inj}}=0, -9,$ and $-23~\mu $A at E3.
In order to compare our data with the predictions of Eqn. (1) we discuss below the measured noise $S$ as a function of $V$.
For $I_{\mathrm{inj}}=0~\mu $A, the observed $S$ is independent of the magnetization configuration, and monotonically increases as $|V|$ increases, as expected from conventional shot noise theory \cite{ref14}.
The finite $S$ at $V=0$ stems from thermal noise.
This curve is perfectly reproduced by the conventional formula given by
\begin{equation}
S=\frac{4k_{\mathrm{B}}T_{e}}{R_{\mathrm{d}}}+2F\left(eI\coth \left(\frac{eV}{2k_{\mathrm{B}}T_{e}}\right)-\frac{4k_{\mathrm{B}}T_{e}}{R_{\mathrm{d}}}\right),
\label{eq2}
\end{equation}
where $R_{\mathrm{d}}$, $T_{e}$, and $F$ are the differential resistance at a given bias, the electron temperature, and the Fano factor, respectively.
For $T_{e}=$1.6 K, the fitting to this formula gives $F=0.78\pm 0.04$ as shown in Fig. \ref{Fig:2}(a) by the dotted curve.
The Fano factor is slightly reduced from unity as expected for a tunneling junction.
This is most probably due to the contribution of defects in the tunneling barrier which are unavoidably created during the (low-temperature) growth process \cite{ref25,ref26,ref27}. 

When $I_{\mathrm{inj}}$ is finite, i.e. for $-9$ and $-23~\mu $A, $S$ starts to depend on the magnetization configuration as expected by Eqn. (1) except for the low bias region, which is dominated by thermal noise (See Fig.1(b) and 3(a)). Moreover, for $I_{\mathrm{inj}}=-23~\mu $A, the horizontal shift of $S$ between P and AP configuration is 0.28 mV, matching closely the magnitude of the non-local voltage signal of $\Delta V=0.30$~mV, shown in Fig. 1(b).
This indicates the validity of the scenario put forward in the introduction and demonstrates that "noise spectroscopy" successfully detects the nonequilibrium spin accumulation $\Delta \mu $ (see also Fig. \ref{Fig:1}(c)).
We emphasize that the observed difference between the two different magnetic configurations is reproducible.
Actually, the $B$-dependence of $S$ was measured for constant $V=-6.8$~mV in an independent measurements, shown in Fig. \ref{Fig:2}(b).
Clear spin-valve-like changes were observed for $I_{\mathrm{inj}}=-9~\mu$A and $-23~\mu$A reflecting magnetization switching in the electrodes, while no detectable change was observed for $I_{\mathrm{inj}}=0~\mu$A.
Thus, the configuration-dependent contribution of the noise measured at finite $I_{\mathrm{inj}}$ is undoubtedly the shot noise associated with spin accumulation. 

In order to distinguish contributions from the voltage drop across the barrier $V$ and the spin accumulation in the channel $\Delta \mu $ in the measured current, we adopt the following model based on the Landauer-B\"{u}ttiker formalism \cite{ref14,ref28}.
Currents for P ($I_{\mathrm{P}}$) and AP ($I_{\mathrm{AP}}$) states are written as
\begin{equation}
I_{\mathrm{P/AP}}=\frac{e}{h}T_{1}\left(eV\pm \Delta \mu/2 \right)+\frac{e}{h}T_{2}\left(eV\mp \Delta \mu/2 \right),
\end{equation}
with $T_{1(2)}$ the sum of all the transmission probabilities of the tunneling channels from the nonmagnetic GaAs conduction band states to the majority (minority) spin states in the ferromagnetic (Ga,Mn)As.
The first and second terms in Eqn. (3) represent tunneling currents with up- and down-spin, respectively.
By using the ratio of the tunneling conductances, $\alpha = T_{1}/(T_{1}+T_{2})$ and $\beta = T_{2}/(T_{1}+T_{2})$, Eqn. (3) can be rewritten as $I_{\mathrm{P/AP}}=\frac{2e}{h}\bar{T}eV \pm \frac{e}{h} \left(\alpha -\beta\right) \bar{T}  \Delta \mu $, where $\bar{T} \equiv \left( T_{1}+T_{2}\right)/2$.
To separate the currents driven by the applied bias $V$ and the spin accumulation $\Delta \mu $ we define the charge current $I_{C}\equiv \frac{2e}{h}\bar{T} eV$ and the spin current $I_{S}\equiv \frac{e}{h}\bar{T} \Delta \mu$.
Note that this spin current is solely driven by the spin accumulation in the $n$-GaAs channel.
These currents can be extracted from experiment using the relations, $\left\langle I_{C}\right\rangle=\frac{I_{\mathrm{P}}+I_{\mathrm{AP}}}{2}$ and $\left\langle I_{S}\right\rangle=\frac{I_{\mathrm{P}}-I_{\mathrm{AP}}}{2\left(\alpha -\beta\right)}$.
By using the experimentally obtained value of $\gamma $, we determined $\alpha $ and $\beta $ as 0.91 and 0.09, respectively.
The energy dependence of these values is negligible in the present experiment at low bias.

We first focus on the excess shot noise in the high bias region ($|eV|>\Delta \mu /2 $, $2k_{B}T$) and discuss the measured noise in parallel configuration $S_{\mathrm{P}}$ and anti-parallel configuration $S_{\mathrm{AP}}$.
Note that, because $|eV|>\Delta \mu /2$, either $eV\pm \Delta \mu /2>0 $ or $eV\pm \Delta \mu /2<0$ holds (see Fig. \ref{Fig:1}(c)).
By similar transformations,, $S_{\mathrm{P}}$ and $S_{\mathrm{AP}}$ can be written as $S_{\mathrm{P/AP}}=\frac{4e^2}{h}\bar{T}FeV \pm \frac{2e^2}{h} \left(\alpha -\beta\right) \bar{T} F \Delta \mu $.
Accordingly, the conventional shot noise $S_{C}$ and excess shot noise $S_{S}$ are defined as $S_{C}\equiv \frac{4e^2}{h}\bar{T} FeV=\frac{S_{\mathrm{P}}+S_{\mathrm{AP}}}{2}$ and $S_{S}\equiv \frac{2e^2}{h}\bar{T} F\Delta \mu=\frac{|S_{\mathrm{P}}-S_{\mathrm{AP}}|}{2\left(\alpha -\beta \right)}$, respectively.

Using experiments as the one shown in Fig. 2a we can experimentally derive the relation between $\left\langle I_{S}\right\rangle$ and $S_{S}$ exactly as in conventional shot noise experiments.
By tuning $I_{\mathrm{inj}}$ and/or by choosing different electrodes for detection (E3, E4, or E5), we perform noise measurements for different values of $\Delta \mu $.
In Fig. \ref{Fig:2}(c), we show the corresponding data $S_{S}$ vs $\left\langle I_{S}\right\rangle$ and $S_{C}$ vs $\left\langle I_{C}\right\rangle$ for $|eV|>\Delta \mu /2$, $2k_{B}T$, for several values of $I_{\mathrm{inj}}$ and for different detection electrodes. 
$S_{C}$ is well fitted by the linear function $S_{C}=2eF|\left\langle I_{C}\right\rangle|$ (see inset in Fig. \ref{Fig:2}(c)), which yields $F=0.77\pm 0.04$, consistent with the zero-injection case discussed above (see Fig. \ref{Fig:2}(a)).
Similarly, we find that the excess shot noise $S_{S}$ also linearly depends on $\left\langle I_{S}\right\rangle$ and, more importantly, satisfies the relation $S_{S}=2eF|\left\langle I_{S}\right\rangle|$ (dashed line) with the same Fano factor obtained for the conventional shot noise.
Both the linear relations and the equal Fano factors justify the applied procedure to extract the excess shot noise which is generated by the spin accumulation.

\begin{figure}[bp]
\center \includegraphics[width=.85\linewidth]{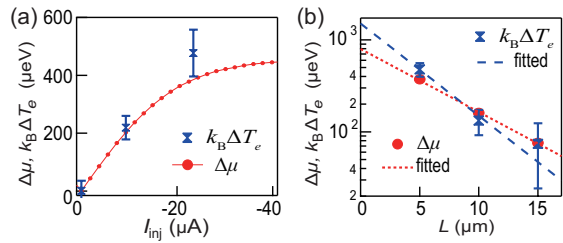}
\caption{(color online). (a) $I_{\mathrm{inj}}$-dependence of the extracted $k_{\mathrm{B}} \Delta T_{e}$ and $\Delta \mu $ at detector E3. (b) $L$-dependence of $k_{\mathrm{B}} \Delta T_{e}$ and $\Delta \mu $ for $I_{\mathrm{inj}}=-23~\mu $A.}
\label{Fig:3}
\end{figure}

In the low bias region ($|eV|\le \Delta \mu /2 $), $S_{\mathrm{P}}$ and $S_{\mathrm{AP}}$ are expected to have opposite slopes, as shown in Fig. 1(b).
For finite temperature, the pronounced kinks at $V=\pm \Delta \mu /2e $ are smeared, and Eqn. (1) need to be replaced by $S_{\uparrow /\downarrow} \propto \alpha |eV\pm \Delta \mu/2|\cosh (\frac{eV\pm \Delta \mu/2}{2k_{\mathrm{B}} T_{e}})+\beta |eV\mp \Delta \mu/2|\cosh (\frac{eV\mp \Delta \mu/2}{2k_{\mathrm{B}} T_{e}})$.
For an electron temperature of $T_{e}=1.6~$K, equal to the bath temperature, the kinks should be detectable \cite{ref9}. However, this structure is not resolved in experiment due to the thermal smearing of the distribution function by the injected hot electrons.

The latter can be seen by estimating the degree of nonequilibrium in terms of the effective electron temperature by fitting  $\left(S_{\mathrm{P}}+S_{\mathrm{AP}}\right)/2$ to Eqn. (2). 
The estimated value of $k_{\mathrm{B}}T_{e}$ is always larger than $\Delta \mu $.
$k_{\mathrm{B}}\Delta T_{e}$ and $\Delta \mu $ are plotted against $I_{\mathrm{inj}}$ and the injector-detector separation $L$ in Figs. \ref{Fig:3}(a) and \ref{Fig:3}(b), respectively, where $\Delta T_{e}$ ($\equiv T_{e}-1.6~$K) is the effective temperature rise.
One can see that, as $\Delta \mu $ increases as a function of $I_{\mathrm{inj}}$, $k_{\mathrm{B}}\Delta T_{e}$ increases similarly.
To the best of our knowledge, direct information such as $k_{\mathrm{B}}\Delta T_{e}$ to characterize the degree of nonequilibrium due to spin accumulation has never been demonstrated so far.
Moreover, as $\Delta \mu $ relaxes according to the diffusion equation, $k_{\mathrm{B}}\Delta T_{e}$ also relaxes.
By fitting $\Delta \mu $ and $k_{\mathrm{B}}\Delta T_{e}$ to $\Delta \mu \propto \exp (-L/\lambda _{S} )$ and $k_{\mathrm{B}}\Delta T_{e}\propto \exp (-L/\lambda _{e})$, we obtain a spin relaxation length $\lambda _{S}=5.6~\mu $m and an energy relaxation length $\lambda _{e}=4.3~\mu $m in the channel, respectively (see the dashed curves in Fig. \ref{Fig:3}(b)).
The good agreement between the two relaxation lengths indicates that spin relaxation is accompanied by energy relaxation.


Finally, we note that $S_{C}$ and $S_{S}$ described here are identical due to the absence of spin flips \cite{ref2,ref4,ref6p5,ref8} or Coulomb interaction effects \cite{ref5p5}, that would otherwise generate different Fano factors for spin and charge current.
By measuring this difference, general spin-dependent shot noise physics can be addressed.
In addition, the present demonstration can lead to a new probe to explore the spin-transfer torque physics \cite{ref7}, the spin heat accumulation \cite{ref29}, and the spin-dependent chiral edge states in topological insulators \cite{ref30}.

In summary, we have shown that a non-equilibrium spin accumulation can be detected in the noise spectrum including the distribution function.
The Landauer-B\"{u}ttiker formalism allowed us to disentangle conventional shot noise associated with the charge current and excess shot noise, connected with a spin current.
However, due to the elevated electron temperature stemming from non-local spin injection, it was not possible to explore excess shot noise in the limit $V\to 0$, i.e. when only a pure spin current is flowing.


\begin{acknowledgments}
We appreciate fruitful comments and supports from M. Ferrier and R. Sakano.
This work is partially supported by Grant-in-Aid for Scientific Research (S) (No. 26220711) from JSPS, Grant-in-Aid for Scientific Research on Innovative Areas "Fluctuation \& Structure" (No. 25103003) from MEXT, Grant-in-Aid for Young Scientists (Start-up) (No. 25887037) from JSPS, the Murata Science Foundation, D.W. and J.N. acknowledge support from German and Japanese Joint Research Program, the German Science Foundation (DFG) via SFB 689, and the Collaborative Research Program of the Institute for Chemical Research, Kyoto University.
\end{acknowledgments}


\renewcommand{\bibname}{}

\part*{Supplemental Material}

\section{Estimation of the spin asymmetry coefficient}
Here, we explain how to estimate the spin asymmetry coefficient $\gamma _{\mathrm{det}}$ at the detection electrode.
The non-local spin-valve signal $\Delta V$ as a function of the channel length is described as [24, 25]
\[
\Delta V=\frac{\gamma _{\mathrm{inj}}\gamma _{\mathrm{det}}I _{\mathrm{inj}}\lambda  _{\mathrm{s}}\rho }{A}\exp \left(-\frac{L}{\lambda  _{\mathrm{s}}}\right),
\]
where $\rho $ and $A$ are the resistivity and the cross-sectional area of the $n$-GaAs cannel, respectively, and $\gamma _{\mathrm{inj}}$ is the spin asymmetry coefficient at the injection electrode.
$\lambda  _{\mathrm{s}}$ and $L$ are, respectively, spin relaxation length and the injector-detector separation as in the main text.
In our device, $\rho =6.7\times 10^{-4}~\Omega $m and $A=50~\mathrm{m}^{2}$.
By using these values and the estimated $\lambda  _{\mathrm{s}}$ value (see main text), we obtain $\gamma _{\mathrm{inj}}\gamma _{\mathrm{det}}$.
Figure 4 shows the obtained $(\gamma _{\mathrm{inj}}\gamma _{\mathrm{det}})^{1/2}$ at E3 as a function of $I _{\mathrm{inj}}$.
The value of $(\gamma _{\mathrm{inj}}\gamma _{\mathrm{det}})^{1/2}$ monotonically increases as $I _{\mathrm{inj}}$ decreases and saturates at very low bias region, fully consistent with the previous work [22].
Assuming $\gamma =\gamma _{\mathrm{det}}=\gamma _{\mathrm{inj}}$ for sufficiently low bias [21, 22], we obtain $\gamma _{\mathrm{det}}=0.82\pm 0.03$, which we use in the main text.
Note that $I _{\mathrm{C}}$ in Fig. 3c of the main text is well below 1 $\mu $A and therefore $\gamma _{\mathrm{det}}$ is constant.
\begin{figure}[bp]
\center \includegraphics[width=.90\linewidth]{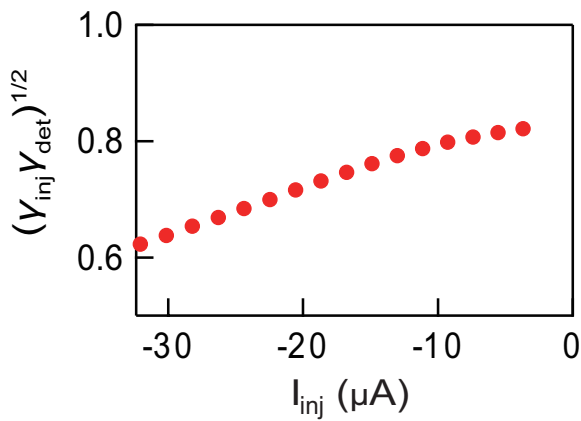}
\caption{Spin asymmetry coefficient. The obtained $(\gamma _{\mathrm{inj}}\gamma _{\mathrm{det}})^{1/2}$ at detector E3 as a function of injection current $I _{\mathrm{inj}}$.}
\label{Fig:S1}
\end{figure}

\section{Differential resistance at the detection barrier}
Here we discuss the differential resistance $dV/dI$ of the tunneling barrier when a spin accumulation is present.
Our tunneling junction shows a zero bias peak in $dV/dI$, which becomes prominent as the temperature lowers.
Figure 5 presents the measured $dV/dI$ at the electrode E3 for $I_{\mathrm{inj}}=0, -9,$ and $-23~\mu $A.
For $I_{\mathrm{inj}}=0$, the peak position is independent of the magnetic configuration, while for finite $I_{\mathrm{inj}}$ the peak positions for P and AP configurations shift to the opposite directions. These shifts are in quantitative agreement with those of the noise signal that are discussed in the main text.
Moreover, the peak gets broader as $I_{\mathrm{inj}}$ increases.
This suggests an increasing electron temperature and is consistent with the behavior of the noise. In the main text, we quantitatively discuss this observation in terms of the effective temperature rise $\Delta T_{e}$.
\begin{figure}[bp]
\center \includegraphics[width=.90\linewidth]{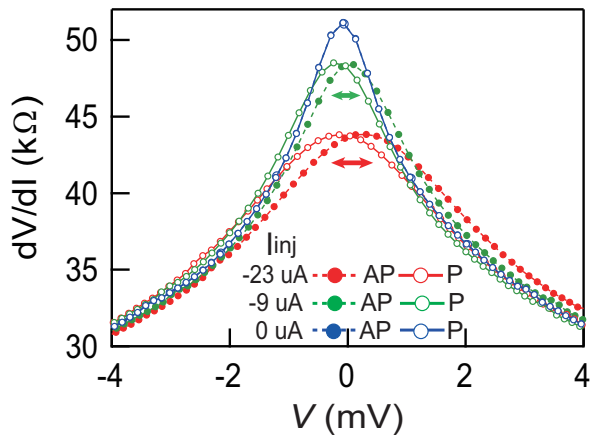}
\caption{Differential resistance in the presence of a spin accumulation. Measured $dV/dI$ at detector E3 as a function of $V$ for P and AP configurations for injection currents $I_{\mathrm{inj}}=0, -9,$ and $-23~\mu $A.}
\label{Fig:S2}
\end{figure}

\section{The result for positive injection current}
In the main text, we present results only for negative injection currents.
Here we present the results for a positive injection current, where the bias direction corresponds to forward biasing of the Esaki diode.
Figure 6 shows the measured noise at E3 for $I_{\mathrm{inj}}=23~\mu $A.
The direction of the horizontal shift of the curves for P and AP alignment is opposite to that found in case of a negative injection current.
This is consistent with the reversal of the accumulated spin species.
We note that the obtained $S_{S}$ is proportional to $I_{S}$, being consistent with the discussion in the main text. 
\begin{figure}[bp]
\center \includegraphics[width=.90\linewidth]{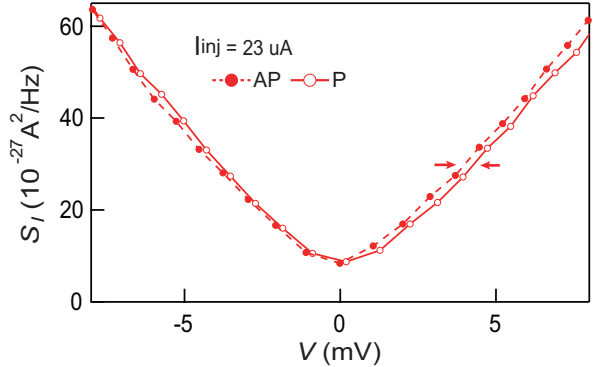}
\caption{Measured noise signal for positive injection current and extracted spin shot noise components. Measured $S_{I}$ at detector E3 as a function of $V$ for P and AP configurations for injection current $I_{\mathrm{inj}}=23~\mu $A.}
\label{Fig:S3}
\end{figure}

\section{Details of noise measurement and the calibration}
Figure 7A shows the detailed set-up of our experiment system.
The DC bias voltage of the detection electrode is measured by the multimeter (Keithley 2000) after amplification to avoid noise from the voltmeter.
A typical procedure of a noise measurement including magnetization is described as follows:
\begin{enumerate}
  \item 	A constant spin injection current $I_{\mathrm{inj}}$ is applied to injector E2.
  \item To saturate the magnetization of injection and detection electrodes, we set $B=2000$ Oe.
  \item Then, we set $B=-100$ Oe and measure the bias dependence of the noise for P configuration, where the cross-correlated voltage spectra, the differential resistance, the bias voltage, and the bias current are recorded for each point (see Fig. 7B).
  \item To carry out the measurement for AP configuration, we set $B=-175$ Oe and switch the magnetic configuration, and then we set $B=-100$ Oe to measure (see Fig. 7B).
\end{enumerate}

Figure 8 shows the equivalent circuit of the detection line for the DC signal.
To measure the current-voltage ($I-V$) characteristic across the detection contact (E3, E4, or E5), we used the constant current method; The bias voltage Vapplied was applied to the detection contact, whose resistance is typically 45 k$\Omega $, through the large resistor R1.
We directly measured both the current I and the voltage drop V across the detection contact.
The measured $I-V$ traces for P and AP configuration are horizontally shifted exactly by the non-local voltage (due to spin accumulation) as shown in the right hand of Fig. 8.
Here, the current difference $I_{\mathrm{P}} - I_{\mathrm{AP}}$ for a given voltage is not the change in charge current but the spin current.
And consequently the difference of the corresponding noise signals, shown in Fig. 3a, is due to spin shot noise.
In order to obtain sufficient resolution of the noise signal, we averaged 15,000 cross-correlation spectra.
The typical time to take one point for the noise signal in Fig. 3a in the main text is about 16 min.
For the magnetic field dependence shown in Fig. 3b, only 9,000 spectra were averaged, thus the noise signals are slightly scattered when compared to the data in Fig. 3a.
The peak around zero magnetic field due to DNP in Fig. 2b and Fig 7B appears within $\pm$~100 Oe as a Lorentz curve (see the gray dashed curve in Fig. 7B) [22].
Outside the DNP peak, the non-local voltage still decays with  increasing external magnetic field but because of a different mechanism.
At the measurement temperature, the  magnitude of the cubic magnetic anisotropy field of (Ga,Mn)As along [100] and that of the uniaxial anisotropy field along [1-10] are comparable and thus, the  total magnetic easy axis lies along somewhere between [100] and [1-10].
When the external magnetic field is swept from zero to either positive or negative direction along the electrode //[1-10], reorientation of the magnetization from the easy axis toward the field occurs.
In spin Esaki diodes, the value of the spin asymmetry coefficient of the tunneling conductance $\lambda $, i.e., spin injection efficiency, depends on the magnetization orientation with respect to the crystallographic direction of GaAs, which we ascribed to the tunneling anisotropic spin polarization (TASP).
Thus, the non-local voltage, a direct measure of spin accumulation, still changes as a function of the magnetic field due to the magnetization reorientation.
Moreover, since we averaged for a long time to take each point of the spectrum (typically 16 min for each point), the influence of the DNP is negligibly reduced.
To estimate the voltage spectral density $S_{V}$, we performed a histogram analysis [27] for the data between 16 kHz and 160 kHz (9,001 points).
As mentioned in the method section, the $S_{V}$ contains not only the intrinsic signal $S_{I}$ but also extrinsic thermal noise from R2 ($S_{I}^{R2}=4k_{\mathrm{B}}T/R2$), the channel resistance R3 ($S_{I}^{R3}=4k_{\mathrm{B}}T/R3$), and the input current noise of the amplifier ($S_{I}^{amp}$).
For example, R3 between the electrodes E3 and E6 is 7.01 k$\Omega $.
The contribution of these current noise sources to $S_{V}$ is given by
\begin{multline*}
\frac{S_{V}}{G^{2} }=\left(\frac{\frac{dV}{dI} \mathrm{R2}}{\frac{dV}{dI}+\mathrm{R2}+\mathrm{R3}} \right)^{2}S_{I} \\+ \left(\frac{\left(\frac{dV}{dI}+\mathrm{R3}\right)\mathrm{R2}}{\frac{dV}{dI}+\mathrm{R2}+\mathrm{R3}}\right)^{2}\left(S_{I}^{\mathrm{R2}}+S_{I}^{\mathrm{amp}}\right)\\+ \left(\frac{\mathrm{R2R3}}{\frac{dV}{dI}+\mathrm{R2}+\mathrm{R3}} \right)^{2} S_{I}^{\mathrm{R3}} ,
\end{multline*}
where $G=102.2$ is the gain of the amplifier after cross-correlation is taken into account.
In our analysis the $S_{I}^{\mathrm{R2}}$, $S_{I}^{\mathrm{amp}}$, and $S_{I}^{\mathrm{R3}}$ contributions are subtracted from the experimentally obtained $S_{V}$ values by using this equation and the measured stage temperature $T$.
\begin{figure}[bp]
\center \includegraphics[width=.90\linewidth]{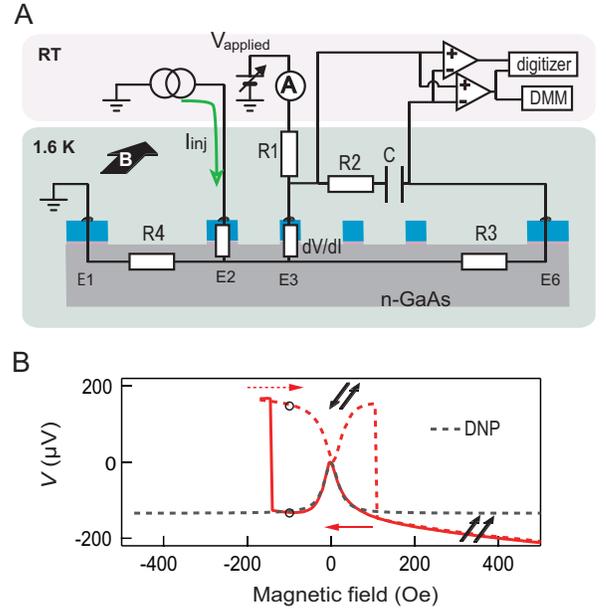}
\caption{Schematic illustration of the experiment and a typical spin accumulation signal. A, Schematic diagram of sample and measurement system. $dV/dI$ and R3 are the differential resistance of the detection tunneling barrier and channel resistance between contacts E3 and E6, respectively. B, Non-local voltage signal detected between E3 and E6 as a function of an in-plane magnetic field $B$ for $I_{\mathrm{inj}}=-23~\mu $A, where the magnetic field is first ramped from 500 to -175 Oe and then from -175 to 500 Oe. The black circles mark the points where noise measurements were done. The gray dashed curve is a fit of the data with the expected Lorentz curve [22].}
\label{Fig:S4}
\end{figure}
\begin{figure}[bp]
\center \includegraphics[width=.90\linewidth]{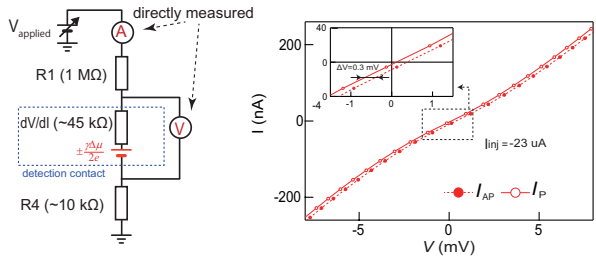}
\caption{Schematic circuit used in experiment and a typical $I-V$ curve. An applied voltage $V_{\mathrm{applied}}$ drives a constant charge current $I_{\mathrm{P}} + I_{\mathrm{AP}}$. Due to spin-charge coupling a non-local voltage $V$ develops across the ferromagnetic contact E3 (indicated by the modified voltage source) which is measured. This voltage is due to the spin accumulation underneath contact E3 (see Fig. 2c) and drives the spin current $I_{\mathrm{P}} - I_{\mathrm{AP}}$, which can be directly read off from the $I-V$ characteristic, displayed on the right hand side.}
\label{Fig:S5}
\end{figure}

\end{document}